\documentclass[showpacs,prl,twocolumn,aps]{revtex4}

\usepackage{amsmath}
\usepackage{amssymb}
\usepackage{latexsym}
\usepackage{amsfonts}
\usepackage{epsfig}
\usepackage{psfrag}
\usepackage{graphicx}

\usepackage{color}

\newcommand{\bea}{\begin{eqnarray}}
\newcommand{\ea}{\end{eqnarray}}
\newcommand{\eea}{\end{eqnarray}}
\newcommand{\ord}{\,{\cal O}}

\begin{document}

\title{Catalysis of Schwinger Vacuum Pair Production} 

\author{Gerald V. Dunne$^{1,2}$, Holger Gies$^{2,3}$, 
and Ralf Sch\"utzhold$^{4}$ }

\affiliation{$^1$ Department of Physics, University of Connecticut, 
Storrs CT 06269, USA
\\
$^2$Theoretisch-Physikalisches Institut, 
Friedrich-Schiller-Universit\"at Jena, 
D-07743 Jena, Germany
\\
$^3$Helmholtz Institute Jena, 
D-07743 Jena, Germany
\\
$^4$ Fakult\"at f\"ur Physik, Universit\"at Duisburg-Essen,
D-47048 Duisburg, Germany}

\begin{abstract}
We propose a new catalysis mechanism for non-perturbative vacuum
electron-positron pair production, by superimposing a plane-wave 
X-ray probe beam with a strongly focused optical laser pulse,  
such as is planned at the Extreme Light Infrastructure (ELI) facility.  
We compute the absorption coefficient arising from vacuum polarization 
effects for photons below threshold in a strong electric field.
This set-up should facilitate the (first) observation of this 
non-perturbative QED effect with planned light sources such as ELI
yielding an envisioned intensity of order $10^{26}\rm W/cm^2$. 
\end{abstract}

%\date{\today}


\maketitle

Soon after Dirac's seminal discovery that a consistent relativistic 
quantum description of electrons entails the existence of positrons, 
pictured as holes in the Dirac sea \cite{Dirac},
it was realized that a strong enough electric 
field can create electron-positron pairs out of the 
vacuum \cite{sauter,eh,schwinger,ringwald,dunne}.
For a constant electric field $E$, the electron-positron pair creation rate
(i.e., the number of pairs per unit time and volume) is given by
  \cite{eh,schwinger}
\bea
\label{Schwinger}
{\mathfrak R}_{e^+e^-}=\frac{e^2E^2}{4\pi^3}
\sum\limits_{n=1}^{\infty}
\frac1{n^2}
\exp\left\{-n\pi\,\frac{m^2}{eE}\right\}
\,,
\ea
where $m$ is the electron rest mass, and $e$ the elementary charge (we use
$\hbar=c=\varepsilon_0=1$).
Apart from the striking experimental possibility to create 
matter out of the vacuum just by applying a strong electric 
field, this quantum field theoretical prediction is of 
fundamental importance due to its purely non-perturbative nature:
the rate (\ref{Schwinger}) has no Taylor expansion in (positive) 
powers of the coupling strength $e$, 
so no Feynman diagram $\propto e^{2n}$ of arbitrarily large (finite) 
order $n$ can describe this phenomenon. 
While non-perturbative tunneling processes are well-studied in
quantum chromo-dynamics (QCD) at both the fundamental and 
phenomenological level \cite{shuryak},
quantum electro-dynamics (QED) offers the possibility of a
  well-controllable direct experimental observation of such non-perturbative vacuum effects
\cite{ritus,mourou}.

The relevant critical field strength scale, 
$E_{\text{cr}}=m^2/e\simeq 1.3\times 10^{18}$V/m, 
is set by the exponent in (\ref{Schwinger}), 
and corresponds to a field intensity 
$I_{\text{cr}}=E_{\text{cr}}^2 \simeq 4.3\times10^{29}$W/cm${}^2$.
For $E\ll E_{\rm cr}$, 
the pair creation rate (\ref{Schwinger}) is  strongly 
(exponentially) suppressed, even for the dominant $n=1$ term.
For instance, an electric field $E$ corresponding to an intensity of 
$I=10^{26}\rm W/cm^2$ yields a Schwinger exponential factor
$\exp\{-\pi m^2/(eE)\}\sim 10^{-90}$, and for $I=10^{27}\rm W/cm^2$
this factor  is  $\sim 5\times10^{-29}$.
It has been argued \cite{bulanov} that for certain focused laser 
beam configurations, a space-time volume enhancement factor 
might render observation feasible for $I=10^{27}\rm W/cm^2$.
Unfortunately, $10^{27}\rm W/cm^2$ is 
extremely difficult to reach, 
and already with $10^{26}\rm W/cm^2$ it is hard to 
see how the exponential suppression of order $10^{-90}$ 
could be compensated by such a four-volume enhancement factor.  
Thus, while $e^+e^-$ pair creation has already been observed in the 
perturbative multi-photon regime \cite{Burke:1997ew} or due to the 
Bethe-Heitler process \cite{Chen:2009}, 
truly non-perturbative QED vacuum effects have so far 
eluded experimental observation.

Partly motivated by these difficulties, we recently proposed a 
{\it dynamically assisted Schwinger mechanism} 
\cite{Schutzhold:2008pz}, showing that it is possible 
to enhance significantly the pair creation rate (\ref{Schwinger}) 
by superimposing a strong but slow electric field with a weak but 
fast electric field, resulting in a decrease of the effective spectral 
gap between the electron states and the Dirac sea. 
Here, we extend this idea into a potentially realistic experimental 
scenario, proposing a new catalysis mechanism for non-perturbative 
vacuum $e^+e^-$ pair production.
The set-up we propose  is a superposition of a plane-wave X-ray 
probe beam with a strongly focused optical laser pulse, 
$\sim 10^{26}\rm W/cm^2$,
as  may soon be available in the Extreme Light Infrastructure (ELI) 
project \cite{eli}. 
This superposition leads to a dramatic enhancement of the expected 
yield of $e^+e^-$ pairs, and brings the vacuum pair production 
effect significantly closer to the observable regime.

A different implementation of the dynamically assisted Schwinger
  mechanism idea has recently been proposed \cite{dipiazza}, wherein a strong
  low-frequency laser and a weak high-frequency laser collide with a
  relativistic nucleus. This configuration effectively lowers the tunneling
  barrier, thereby increasing the pair production rate. Strong Coulomb fields
  from accelerated ions in combination with lasers can also lead to
noticably higher production rates  in the 
multiphoton regime \cite{Muller:2008ys}.

By contrast, our proposal of a ``catalyzed Schwinger mechanism" can be
  realized with all-optics components, explicitly avoids any above-threshold
  scales, and fully preserves the nonperturbative character of the Schwinger
  mechanism.
For this, we assume that the optical laser pulse is focused to yield a maximum
electric field $E$ such that the magnetic field near the focal point can be
neglected (e.g., by colliding two counter-propagating pulses, such that
  $E=2E_{\text{laser}}$).
Furthermore, since the Schwinger mechanism is dominated by the 
region with the highest field strength $E$, and the X-ray 
wavelength is much smaller than the (optical) focus size, 
we can approximate the optical laser pulse by a constant 
\cite{Ruf:2009zz} 
electric field $E$, in which the X-ray photons propagate. 
And with a large number of coherent photons in 
the intense pulse, we can treat the strong field as being classical.
To compute the expected number of pairs produced, we need  the 
absorption coefficient $\kappa$ of an X-ray photon propagating 
in an electric field, which is encoded in the imaginary
part of the polarization tensor $\Pi_{\mu\nu}$:
\begin{equation}
\kappa_{\|,\bot}= -\frac{1}{\omega}\, \Im\left(\Pi_{\|,\bot}\right) 
\,.
\label{kappa}
\end{equation}
Here $\omega$ is the photon frequency and ${\|, \bot}$ denote the
polarizations of the X-ray photons with respect to the electric field
direction.  The quantities $\Pi_{\|,\bot}$ are eigenvalues of the polarization
tensor $\Pi_{\mu\nu}$ corresponding to the eigendirections which reduce to the
two physical transversal modes in the free-field case. For weak fields $eE/m^2
\ll 1$, photons in a field propagate essentially on the light cone with small
corrections, i.e., the phase velocity is $v\equiv \omega/|\mathbf{k}| = 1 +
\mathcal O (\alpha (eE)^2/m^4)$. Even at stronger fields $eE \sim m^2$, the
phase velocity deviates from 1 only by $\mathcal O(\alpha/\pi)$ corrections
\cite{dittrichgies}. Since $\Pi_{\mu\nu}$ itself is of order $\alpha$,
we can safely set $k^2=\mathbf{k}^2-\omega^2\to0$ inside $\Pi_{\mu\nu}$ for
photons propagating in a laboratory electric field.

Given an initial photon amplitude $A_{\text{in}}$, the outgoing amplitude
after traversing an electric field $E$ of length $L$ is
$A_{\text{out}}={\rm e}^{-\kappa L/2} A_{\text{in}}$.
The survival probability of the photon is $P =
|A_{\text{out}}|^2/|A_{\text{in}}|^2 = {\rm e}^{-\kappa L}$.
Thus, given the rate $n_{\text{in}}$ of incoming photons, the pair creation
rate (i.e., number of pairs per unit time) is (for $\kappa L\ll 1$)
\begin{equation}
n_{e^+e^-} = n_{\text{in}} (1-P) = n_{\text{in}} (1-{\rm e}^{-\kappa L}) 
\simeq \kappa L \, n_{\text{in}},
\label{number}
\end{equation}
where we have ignored multiple-pair production which can occur at higher
order.  
Thus, the technical part of the derivation involves computing the
absorption coefficient $\kappa$. 
The imaginary part of the polarization tensor is \cite{dittrichgies}
\begin{equation}
\Im\left(\Pi_{\|, \bot}\right)
=\frac{\alpha}{8\pi i}
\int\limits_{-1}^{+1} \! d\nu 
\!\int\limits_{{\cal C}} \frac{ds}{\sinh s}\, 
{\rm e}^{-i\varphi(s, \nu)}
N_{\|, \bot}(s, \nu),
\label{integral}
\end{equation}
where the contour ${\cal C}$ is just below the real $s$ axis: ${\cal
  C}={\mathbb R}-i\varepsilon$.
In the relevant physical limit, the functions
$N_{\|,\bot}(s, \nu)$ and $\varphi(s, \nu)$ are analytic apart from
simple poles on the imaginary axis at $s\in i\pi\mathbb N$:
\begin{eqnarray}
\varphi
&=&\frac{m^2s}{eE} \left[1-\frac{\tilde{\omega}^2}{m^2}  
\left( \frac{\cosh s - \cosh
    \nu s}{2 s \sinh s} - \frac{(1-\nu^2)}{4} \right)\right]
%\label{functions}
\nonumber\\
N_\|
&=& \tilde{\omega}^2\left( \cosh \nu s - \frac{\nu \sinh \nu s}{ 
\tanh s} - 2 \frac{(\cosh s - \cosh \nu s)}{\sinh^2 s}\right) \nonumber\\
N_\bot
&=& \tilde{\omega}^2 \left( (1-\nu^2) \cosh s- \cosh \nu s + 
\nu \sinh \nu s \coth
  s \right), \nonumber
\end{eqnarray}
where $\tilde{\omega}^2\equiv \omega^2 \sin^2\theta$, with $\theta$ 
being the angle between the $E$ field and the propagation direction, 
$k_z=|{\bf k}|\cos\theta$.  
Notice that the dependence on $\tilde{\omega}^2=\omega^2\sin^2\theta$ 
arises from the relativistic invariants 
$F_{\mu\nu}k^\nu\to\tilde{\omega}^2$, and also 
$\tilde F_{\mu\nu}k^\nu\to\tilde{\omega}^2$ 
(note that $k_\mu k^\mu\approx 0$).

The presence of these poles at $s\in i\pi\mathbb N$ is an important 
ingredient of Schwinger pair production.
The poles are closely related to instantons of the Euclidean theory 
which describe the tunneling of Dirac-sea electrons to the real
continuum  \cite{kimpage}. 
For $\omega\to0$, the sum over the poles at $s=in\pi$ precisely 
corresponds to the terms in Eq.~(\ref{Schwinger}).
If we replace the electric field $E$ by a magnetic field $B$, the poles 
would lie on the real axis instead and there would be no pair creation 
for all frequencies below threshold $\tilde{\omega}<2m$. 
Above threshold $\tilde{\omega}>2m$, however, pair creation mechanisms  
different from the Schwinger effect set in -- that is why the X-ray 
photons should be below threshold in order to avoid these competing 
effects, as the X-ray may also pass through a region (beside the
focus) where the $B$ field dominates.

In the static
limit, $eE\ll m^2$ and $\tilde{\omega}\to0$, the 
first pole dominates, i.e. the $n=1$ term in Eq.~(\ref{Schwinger}).
However, for larger frequencies 
$\tilde{\omega}=\ord(m)$, implying also $eE\ll \tilde{\omega}^2$, 
this is no longer correct and we have to sum over all poles in the 
lower complex half plane.
In addition to a full numerical evaluation (see below), this resummation
  can approximately be
accomplished via the 
saddle point (stationary phase) technique for 
evaluating the integrals in (\ref{integral}), with $eE/m^2$ 
being the small expansion parameter. 
Starting with the $\nu$ integration, we get three saddle points 
(where $\partial\varphi/\partial\nu=0$) at $\nu_*=\pm 1$ and $\nu_*=0$.
However, the critical points $\nu_*=\pm 1$ produce exponentially 
suppressed contributions, such that the dominant saddle point
is given by $\nu_*=0$.
Since $\nu$ governs the asymmetry between the created electron 
and positron, this is consistent with phase space arguments. 
As a result, the $\nu$ integration produces the phase 
$\varphi(s,\nu_*=0)=s\left(1+\tilde{\omega}^2/(4m^2)\right)-\tanh(s/2)
\tilde{\omega}^2/(2m^2)$, and the pre-factor
$\sqrt{2\pi/|\partial^2\varphi/\partial\nu^2(s,\nu_*=0)|}$, with 
$|\partial^2\varphi/\partial\nu^2(s,\nu_*=0)|=
(\tilde{\omega}^2/2eE)
\left|s\left(s/\sinh s-1\right)\right|
$.
The remaining $s$ integral can also be done by the saddle point 
expansion, deforming the contour ${\mathcal C}$ into the 
lower half-plane $\Im(s)<0$, with dominant contribution from the 
saddle point
$
%\frac{d}{ds}
\partial \varphi(s, \nu=0)/\partial s=0
\Rightarrow s_*=-2i\,{\rm arctan}(2m/\tilde{\omega})
$.
For small $\tilde{\omega}$, the saddle point $s_*$ approaches $-i\pi$, 
while near threshold where $\tilde{\omega}\to 2m$, 
we have $s_*\to -i\pi/2$. 
The exponential factor in $\Im \left(\Pi\right)$ is
 \begin{eqnarray}
&& \exp\{-i\varphi(s_*,\nu_*)\}\nonumber\\
&&=
 \exp\left\{
-\frac{m^2}{e E}
\left[
2\left(1+\frac{\tilde{\omega}^2}{4m^2}
\right)
\,{\rm arctan}\left(\frac{2m}{\tilde{\omega}}\right)
-\frac{\tilde{\omega}}{m}
\right]\right\} \nonumber\\
&&\sim
 \begin{cases}
 \exp\left\{-\frac{m^2}{eE}\pi\right\}\quad,\quad  \tilde{\omega}\to 0\cr
 \exp\left\{-\frac{m^2}{eE}(\pi-2)\right\}\quad,\quad  \tilde{\omega}\to 2m
 \end{cases}.
 \label{enhancement}
\ea
At small $\tilde{\omega}$, we recognize the familiar constant-field 
exponential suppression factor, $\exp\{-m^2\pi/(eE)\}$ from (\ref{Schwinger}).
But as $\tilde{\omega}$ approaches the threshold value 
$\tilde{\omega}\to 2m$, there is a dramatic enhancement, 
by a factor $\exp\{2m^2/(eE)\}$, which is very large in the 
$eE\ll m^2$ regime. 
This exponential enhancement is the key feature of our catalysis proposal. 
In spite of the exponential enhancement, our result is still purely
non-perturbative and thus qualitatively different from all 
perturbative multi-photon effects $\propto(eE)^n$. 
The threshold value of the enhancement factor can also be obtained 
from an interaction picture analysis of the appropriate matrix element
\cite{preparation}. 

To obtain precise estimates, we need also the prefactors, 
not just the exponential factor. 
For general $\tilde{\omega}$ below threshold, the $s$ integration 
yields a prefactor
$\sqrt{2\pi/|\partial^2\varphi/\partial s^2(s_*,\nu_*)|}$, with 
$|\partial^2\varphi/\partial s^2(s_*,\nu_*)|
%\left|\frac{d^2\varphi}{ds^2}(s,\nu_*=0)\right|_{s_*}
=
(m/2eE)
(1+4m^2/\tilde{\omega}^2)
$.
%\bea
%\left|\frac{d^2\varphi}{ds^2}(s,\nu_*=0)\right|_{s_*}
%=
%\frac{\tilde{\omega}^2}{2eE}
%\frac{1+4m^2/\tilde{\omega}^2}{\tilde{\omega}/m}
%\,.
%\ea
%
\begin{figure}[t]
\includegraphics[width=0.45\textwidth]{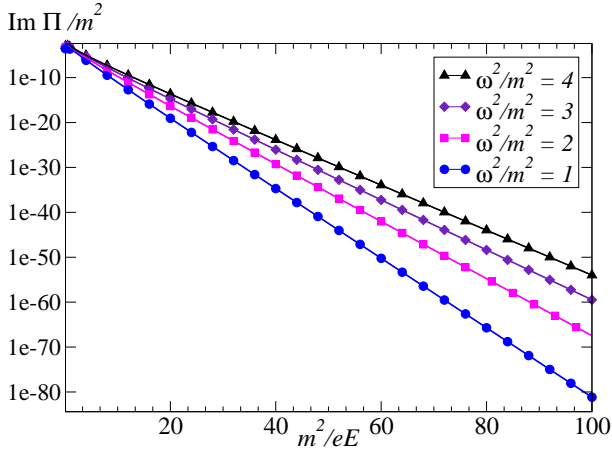} 
\caption{Comparison of numerical evaluation of $\Im (\Pi_\|)$  
[symbols]  with the saddle point expression (\ref{saddle}) [solid lines].}
\label{fig1}
\end{figure}
Combining both prefactors, the exponential factor 
${\rm e}^{-i\varphi(s_*, \nu_*)}$, and $N(s_*, \nu_*)/\sinh s_*$, 
and the appropriate phase, we obtain 
\bea
&&\Im(\Pi_{\|, \bot}) 
\approx 
\frac{\alpha}{8}\,\frac{eE}{m^2}\, 
\left|N_{\|,\bot}\right|\,
\frac{\sqrt{m/\tilde{\omega}}\,\sqrt{1+4m^2/\tilde{\omega}^2}}
{\sqrt{\left|s_*\left(s_*/\sinh s_*-1\right)\right|}} 
\label{saddle}
\\
&&\quad \times \exp\left\{
-\frac{m^2}{e E} \left[ 2\left(1+\frac{\tilde{\omega}^2}{4m^2} \right) \,
{\rm arctan}\left(\frac{2m}{\tilde{\omega}}\right) -\frac{\tilde{\omega}}{m}
  \right] \right\}, \nonumber
\ea
where
\bea
N_\|(s_*,\nu_*)
&=&
\tilde{\omega}^2\left(1-2\frac{\cosh s_*-1}{\sinh^2s_*}\right)
=-4m^2,\\
N_\perp(s_*,\nu_*)
&=&
\tilde{\omega}^2\left(\cosh s_*-1\right)
=
-\frac{8\tilde{\omega}^2 m^2}{\tilde{\omega}^2+4m^2}.
\ea
Expression (\ref{saddle}) is our main technical result.
Notice that at threshold, $\tilde{\omega}=2m$, the difference 
between parallel and perpendicular X-ray polarization disappears 
and (\ref{saddle}) simplifies to 
\bea
\Im(\Pi_{\|, \bot})
\approx
\frac{\alpha}{\sqrt{\pi(\pi-2)}}\,eE\,
\exp\left\{-\frac{m^2}{eE}(\pi-2)\right\}
\,.
\ea
We have also compared the saddle-point expression (\ref{saddle}) with
a direct numerical integration of the double integrals in
Eq.~(\ref{integral}). 
A crucial ingredient for the numerical integration consists in a suitable 
choice of the $s$ contour. 
A parabolic shape $s=-i c_1 + t - i c_2 t^2$ with $0<c_1\leq \pi/2$ and 
$c_2>0$ turns out to be convenient for standard integrators; 
we have typically used $c_1=\pi/2$ and $c_2=3$. 
In the relevant physical regime where $eE\ll m^2$, the agreement 
is excellent, as shown in Fig. \ref{fig1}.

We now discuss the physical implications of the result (\ref{saddle}) 
for the expected yield of electron-positron pairs, using parameters 
relevant to the planned ELI optical laser configuration \cite{eli}. 
\begin{figure}[t]
\includegraphics[width=0.48\textwidth]{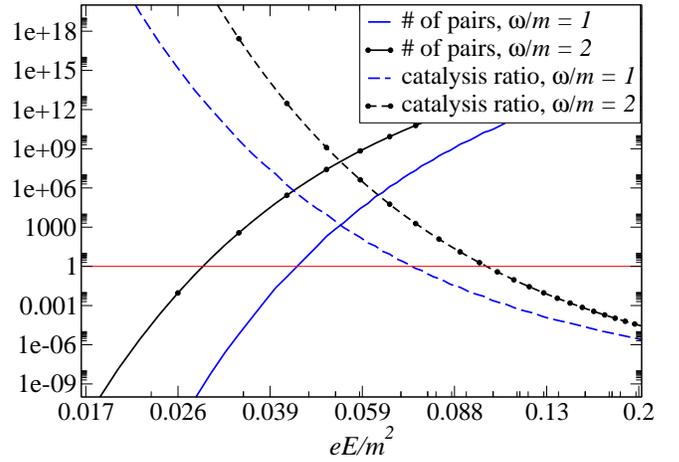} 
\caption{Discovery potential for pair production: number of calatyzed pairs
(solid lines) and the ratio of catalyzed pairs to pairs produced by the
standard Schwinger mechanism (dashed lines) both as a function of the 
field strength. 
Further parameters: $n_{\text{in}}=10^{10}$/pulse, $L=1~\mu$m,
$\theta=\pi/2$, one day of operation at a 1~Hz repetition rate. }
\label{fig2}
\end{figure}
The yield is most sensitive to the parameter $eE/m^2$, the ratio of the 
peak electric field $E$ of the optical laser to the Schwinger critical 
field $E_{\text{cr}}$. 
The envisaged peak intensity at ELI is around $10^{25}-10^{26}{\rm W/cm^2}$, 
so we consider $eE/m^2$ of the order of $1/100$ to $1/10$. 
As the largest possible intensity is reached by maximum focussing in time 
and space, we assume that the temporal and spatial extent of the pulse is 
near $L=1~\mu$m (diffraction limit). 
Pulse synchronization can be achieved by generating the catalyzing X-ray 
beam also from ELI by incoherent Thomson back-scattering of optical photons 
off an laser-accelerated electron bunch. 
We expect a rate of $n_{\text{in}}=10^{10}$ X-ray photons per pulse to be 
realistic. 
In Fig.~\ref{fig2}, the solid lines show the number of pairs produced 
within 86400 pulses, corresponding to one day of operation at a repetition 
rate of 1~Hz (the latter may be realizable, if ELI is built with
diode-pumped amplifiers). 
If the X-ray pulse is at perpendicular incidence $\theta=\pi/2$ and close to
threshold, $\omega\simeq 2m$, our X-ray catalysis pair production sets in at
$eE/m^2>0.029$. With $E=2E_\text{laser}$, this corresponds to a
 laser intensity of $I\simeq 9\times 10^{25}$W/cm${}^2$.
By contrast, standard Schwinger pair production (\ref{Schwinger})
for the same parameters but without an X-ray catalysis is more than 
twenty orders of magnitude smaller.  
This is expressed by the dashed curves in Fig.~\ref{fig2}, exhibiting the 
ratio of the catalyzed pair production to the standard Schwinger production. 
For $eE/m^2>0.101$, i.e., $I>1.1\times 10^{27}$, the standard Schwinger
mechanism starts to dominate, since it scales with $L^4$ as compared to 
the scaling (\ref{number}) of the X-ray catalysis pair production.  
In fact, beyond this field strength further corrections arise, e.g., from
back-reaction and multiple-pair production, but these can be neglected 
for the present discussion since this is beyond the 
envisaged next-generation experimental regime. Thus, we
conclude that our catalyzed pair production mechanism can reduce by more than 
an order of magnitude the required laser 
intensity, from $I>10^{27}$ to
$I\simeq 9\times 10^{25}$W/cm${}^2$, and also yields a drastically enhanced 
signal in the window up to $I<1.1\times 10^{27}$W/cm${}^2$. 
We stress that this {\it exponential} enhancement is a very different mechanism from the {\it linear} volume factor enhancement proposed in \cite{bulanov} based on the pure Schwinger mechanism with a modified effective volume due to spatial focussing, and with very different laser pulse parameters from those at ELI.

Apart from potentially aiding the  first observation of the Schwinger effect, 
which has 
posed a long-standing challenge, these theoretical and experimental 
investigations deepen our understanding of non-perturbative aspects 
of quantum field theory in general.
For example, there is an interesting analogy with the seminal work of Voloshin 
and Selivanov \cite{voloshin} on the phenomenon of induced decay of a 
metastable vacuum (see also \cite{monin}). 
There, they considered processes whereby the decay of a metastable vacuum 
can be induced by the presence of another (massive) particle, which serves 
as a center for the nucleation process, leading to an exponential 
enhancement of the decay probability. 
In our scenario, the massless X-ray photon catalyzes, or induces, 
the vacuum decay process from the strong optical laser field. 
The prefactor contributions in \cite{voloshin} and (\ref{saddle}), which are 
crucial for precise estimates 
of the pair-production yield, are different, but the exponential factors 
are universal. 
In the limit $eE\ll m^2$, and away from threshold, $\tilde{\omega}\ll m$, we can evaluate 
$\Im(\Pi)$ in (\ref{integral}) from the poles of the $1/\sinh (s)$ function, 
and we find the dominant contribution
\bea
\Im(\Pi_{\|})\sim -2\alpha m^2 I_1^2(m\tilde{\omega}/eE)e^{-m^2\pi/eE}\quad ,
\ea
where $I_1$ is the modified Bessel function.
This result has recently been found also in the metastable decay picture \cite{monin2}.

To conclude, our proposal of a catalyzed Schwinger mechanism 
on the one hand introduces a strong amplification mechanism for pair 
production by a tunnel barrier suppression, but on the other hand fully 
preserves the nonperturbative character of the Schwinger mechanism. 
For the latter, we use purely electric field components instead of 
on-shell laser fields and explicitly avoid the presence of any 
above-threshold scales which would open up phase space for perturbative  
production schemes.
\smallskip

RS and HG acknowledge support by the DFG under grants SCHU~1557/1-2,3 
(Emmy-Noether program), SFB-TR12, GI~328/4-1 (Heisenberg program), 
and SFB-TR18,
and GD acknowledges the DOE  grant DE-FG02-92ER40716, 
and  the DFG  grant GK 1523.  
We thank M. Voloshin, A. Monin and especially T. Tajima for 
interesting discussions and helpful correspondence.

\end{document}